\documentclass[submission, Phys]{SciPost}

\hypersetup{
    colorlinks,
    linkcolor={red!50!black},
    citecolor={blue!50!black},
    urlcolor={blue!80!black}
}

\usepackage[bitstream-charter]{mathdesign}
\DeclareSymbolFont{usualmathcal}{OMS}{cmsy}{m}{n}
\DeclareSymbolFontAlphabet{\mathcal}{usualmathcal}

\usepackage[section]{placeins}
\usepackage{lineno}

\begin{document}

\begin{center}{\Large \textbf{
Lepton flavor violation at FCC-ee and CEPC\\
}}\end{center}

\begin{center}
P. Munbodh\textsuperscript{1$\star$}
\end{center}

\begin{center}
{\bf 1} University of California Santa Cruz
\\
* pmunbodh@ucsc.edu
\end{center}

\begin{center}
\today
\end{center}

\definecolor{palegray}{gray}{0.95}
\begin{center}
\colorbox{palegray}{
  \begin{tabular}{rr}
  \begin{minipage}{0.1\textwidth}
    \includegraphics[width=30mm]{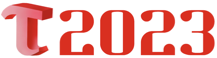}
  \end{minipage}
  &
  \begin{minipage}{0.81\textwidth}
    \begin{center}
    {\it The 17th International Workshop on Tau Lepton Physics}\\
    {\it Louisville, USA, 4-8 December 2023} \\
    \doi{10.21468/SciPostPhysProc.?}\\
    \end{center}
  \end{minipage}
\end{tabular}
}
\end{center}

\section*{Abstract}
{\bf
Lepton flavor violating processes are highly suppressed in the Standard Model. Therefore, if observed, lepton flavor violation would be a clear indication of new physics beyond the Standard Model. We study the process $e^+e^-\to\tau\mu$ at tree-level in the Standard Model Effective Field Theory both on the $Z$-pole and at higher center-of-mass energies. We show that the constraints derived from the future circular $e^+e^-$ colliders FCC-ee and CEPC, are complementary to those obtained from low-energy tau decays at BaBar and Belle, as well as projections from Belle-II. 
}

\section{Introduction}
\label{sec:intro}
Lepton Flavor Violation (LFV) in the Standard Model (SM) is very far beyond current experimental sensitivities as it does not occur at tree-level and is, in addition, suppressed by ratios of neutrino masses to the $W$-boson mass\cite{Marciano:1977wx, Petcov:1976ff}. However, heavy new physics above the TeV scale, which we parametrize by the Standard Model Effective Field Theory (SMEFT) \cite{Grzadkowski:2010es}, might induce larger lepton flavor violating processes that could be searched for. 

Here, we focus on those SMEFT operators that contribute to $e^+e^- \to \tau \mu$ at tree-level. Large Electron-Positron Collider (LEP) analyses provide constraints on $Z\to\tau\mu$ both resonantly on the $Z$-pole and non-resonantly at higher center-of-mass energies $\sqrt{s}\sim 200$ GeV \cite{DELPHI:1992pgs, ALEPH:1991qhf, L3:1993dbo, OPAL:1990fxo, OPAL:1995grn, OPAL:2001qhh}. Since the Future Circular Collider \cite{FCC:2018evy, Bernardi:2022hny} (FCC-ee) and Circular Electron Positron Collider \cite{CEPCStudyGroup:2018ghi, CEPCPhysicsStudyGroup:2022uwl, CEPCStudyGroup:2023quu} (CEPC) will run at higher $\sqrt{s}$ (in addition to the $Z$-pole) and with larger luminosities, they will probe the LFV SMEFT operators much more sensitively than the LEP. In Section \ref{sec:another}, we will discuss the sensitivity projections and compare them to those obtained from low-energy tau decays.

\section{Observables}
For detailed discussion and calculations, see Ref.\cite{Altmannshofer:2023tsa}. There are three classes of operators that contribute to $e^+e^-\to\tau\mu$ at tree-level namely dipole operators, Higgs current operators and four-fermion contact interactions. The total cross-section is given by
\begin{equation}
\label{eq:SMEFT_cross_sec}
\sigma_\text{tot} = \sigma(e^+e^- \to \tau^+ \mu^-) + \sigma(e^+e^- \to \mu^+ \tau^-) = \frac{1}{24\pi} \frac{m_Z^2}{\Lambda^4} \Big( 2(I_0 + \bar I_0) + I_2 + \bar I_2 \Big) ~,
\end{equation}
where $\Lambda$ is the UV scale of new physics, $m_Z$ is the mass of the $Z$, and expressions for the coefficients $I_{0(2)} + \bar{I}_{0(2)}$ are given in Appendix \ref{app:coeff}.

The characteristic scaling of the cross-section $\sigma(e^+e^-\to\tau\mu)$ for each class with the center-of-mass energy can be determined from dimensional analysis. In particular, at large $s$, four-fermion contact interactions give rise to a contribution that is proportional to $s$ whereas the contributions of dipole operators is constant while those of the Higgs current operators is proportional to $1/s$. Finally, operators which contribute to diagrams with an off-shell $Z$ boson experience resonant enhancement in their cross-section on the $Z$-pole. Hence, should new LFV physics reside in the $\tau\mu$ sector, probing these operators at different $\sqrt{s}$ could potentially allow the contributions of each class of operators to be disentangled from one another. Additionally, observables sensitive to the angular distributions and CP asymmetries could allow us to distinguish between the different chirality structures of these operators. 

\begin{figure}[tbh]
\centering
\includegraphics[width=10cm,height=10cm]{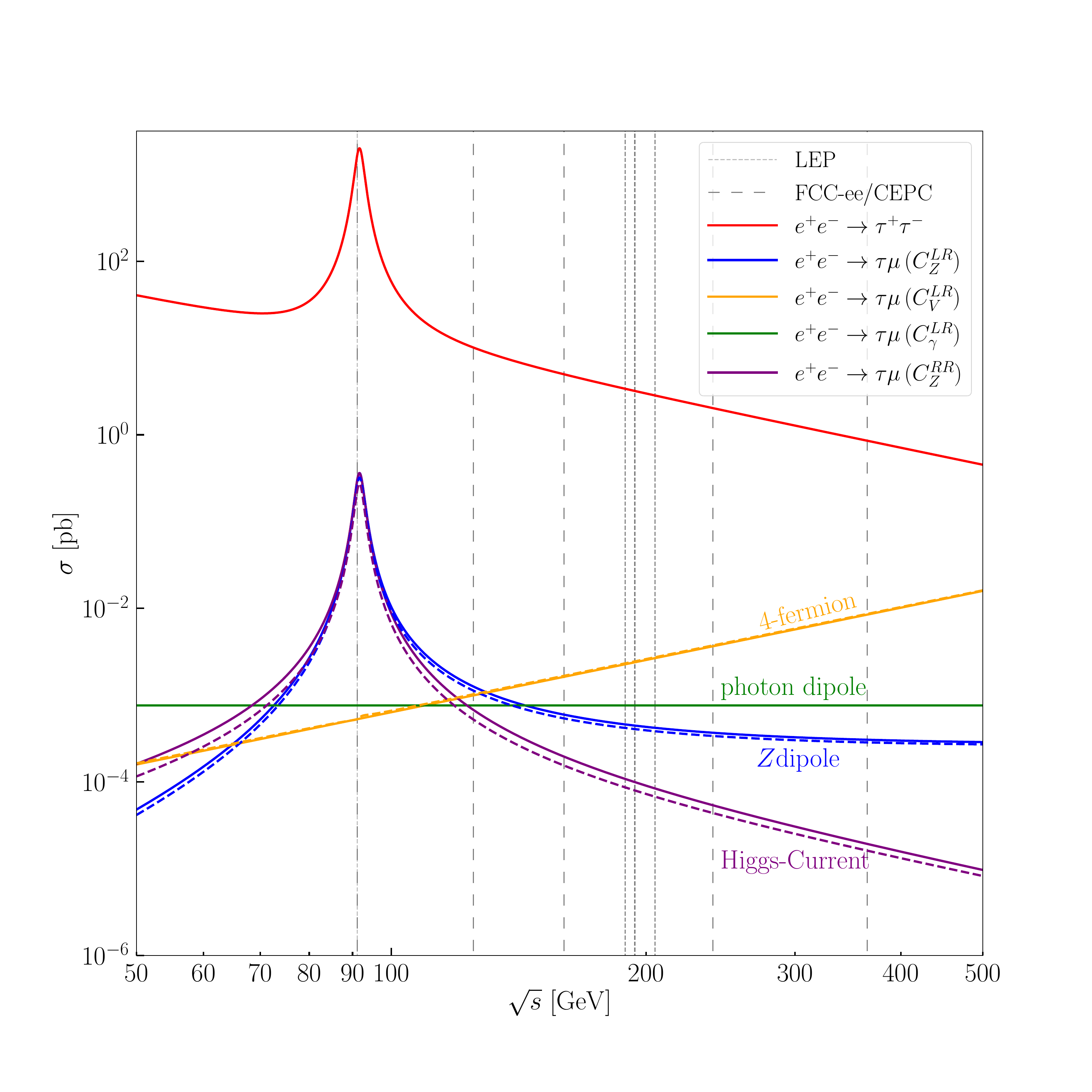}
\caption{The dependence of the cross-section $\sigma(e^+e^-\to\tau\mu)$ on the center-of-mass energy (solid colored lines) for each class of operator setting the Wilson coefficient to 1. The dashed coloured lines include the effects of RGE running~\cite{Jenkins:2013wua, Alonso:2013hga} of the coefficients from a UV scale of $\Lambda=3$ TeV on $\sigma(e^+e^-\to\tau\mu)$. The red line shows the cross-section for the dominant source of background $\sigma(e^+e^- \to \tau^+\tau^-)$.}
\label{fig:cross_sec}
\end{figure}

\section{Expected Sensitivities at FCC-ee and CEPC}
\label{sec:another}
To curb background from $e^+e^- \to \mu^+\mu^-$ where a muon is misidentified as a tau, we require that the tau decays to at least two pions i.e. $\tau_{\rm had} \to \rho \nu \to 2\pi\nu$, $\tau_{\rm had} \to 3\pi\nu$ or $\tau_{\rm had}\to4\pi\nu$, as outlined in Ref.~\cite{Dam:2018rfz}. All remaining backgrounds can be tamed by imposing a cut on 
\begin{equation}
    x = \frac{p_\mu}{p_{\rm beam}}~,
\end{equation}
where $p_\mu$ is the absolute value of the muon momentum and $p_{\rm beam}$ is the beam momentum ~\cite{Dam:2018rfz}. Thus understanding the kinematics of the various background processes is essential. For the background process (i) $e^+ e^- \to W^+ W^- \to \tau_\text{had} \nu \mu \nu$ with two on-shell $W$\footnote{We have verified using MadGraph5 aMC@NLO~\cite{Alwall:2014hca} that the background is still negligible if the $W$ bosons are off-shell.}, kinematically $x<1$ whereas for (ii) $e^+e^-\to\tau^+\tau^-\to\tau_{\rm had}\mu\nu\nu$, the endpoint of the muon momentum is at $x=1$. Finite beam energy spread and detector momentum resolution shifts this endpoint to be slightly above 1. Therefore, selecting only events having $x>1$ will substantially reduce background in the second case whilst almost completely eliminating background in the first case due to the massive $W$ bosons. Furthermore, higher-order background processes such as $e^+ e^- \to W^+ W^-/ZZ \to \tau^+ \nu \tau^- \nu$ also become negligible after applying this cut. On the other hand, the same cut of $x>1$ preserves around half of the signal since the latter is a gaussian centered on $x=1$ (with width dictated by finite detector effects and collision energy spread).

The number of expected signal and background events is given by
\begin{align}
    N_\text{sig} &= \sigma(e^+e^- \to \tau \mu) \times \text{BR}(\tau \to \text{pions}+\nu) \times \mathcal L_\text{int} \times \epsilon^{x_c}_\text{sig} \times \epsilon ~, \\
\label{eq:Nbkg}
 N_\text{bkg} &= \sigma(e^+e^- \to \tau^+ \tau^-) \times 2\times \text{BR}(\tau \to \text{pions}+\nu) \times \text{BR}(\tau \to \mu\bar\nu\nu) \times \mathcal L_\text{int} \times \epsilon^{x_c}_\text{bkg} \times \epsilon  ~,
\end{align}
where $\mathcal{L}_{\rm int}$ is the integrated luminosity, $\epsilon_{\rm sig/bkg}^{x_c}$ are the signal or background efficiencies dependent on the cut $x>x_c$, and $\epsilon \simeq 25\%$ is an estimate of the analysis efficiency included following Ref.\cite{Dam:2018rfz}. Although the momentum cut could be further optimized to gain small increases in the sensitivity, we take $x_c=1$ to illustrate our results. We have further neglected any systematic sources of uncertainty which could become relevant on the $Z$-pole where the number of background events is larger. The effects of initial state radiation have been neglected as well since it only leads to a slight shift in the momentum distribution near $x_c$. The center-of-mass energies, luminosities, beam energy spreads and detector momentum resolutions can be found in Refs.~\cite{FCC:2018evy, Blondel:2019jmp, dEnterria:2021xij, Bernardi:2022hny} for the FCC-ee and in Refs.~\cite{CEPCStudyGroup:2018rmc, CEPCStudyGroup:2018ghi, Gao:2022lew, CEPCPhysicsStudyGroup:2022uwl} for the CEPC.

To estimate the sensitivity of these future machines to the LFV cross-section, we adopt the following criterion
\begin{equation}
    N_{\rm sig} \geq 2 \sqrt{N_{\rm bkg} + N_{\rm sig}}~.
\end{equation}

\begin{figure}[t!]
\centering
\includegraphics[width=0.46\linewidth]{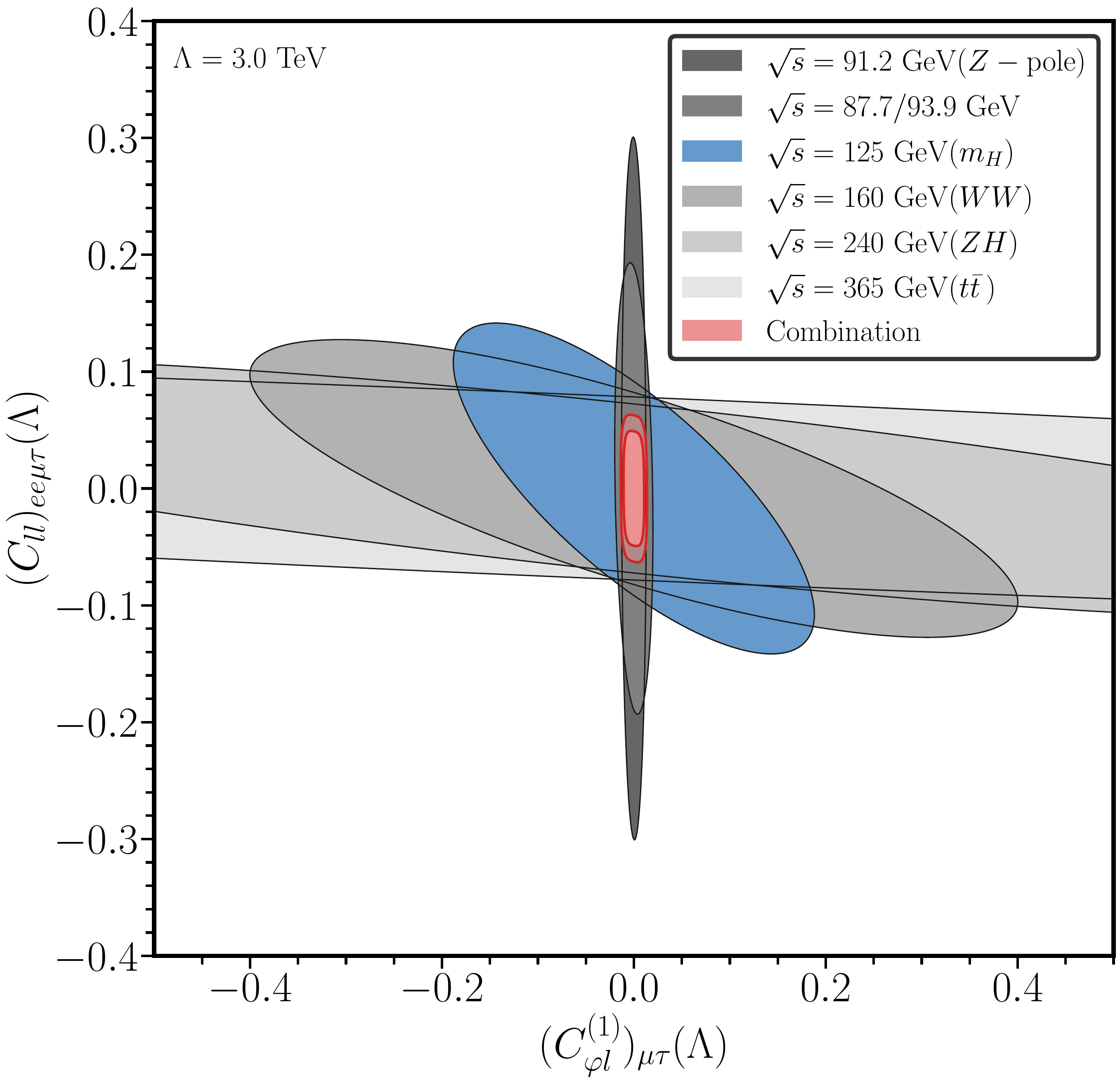} ~~~~
\includegraphics[width=0.46\linewidth]{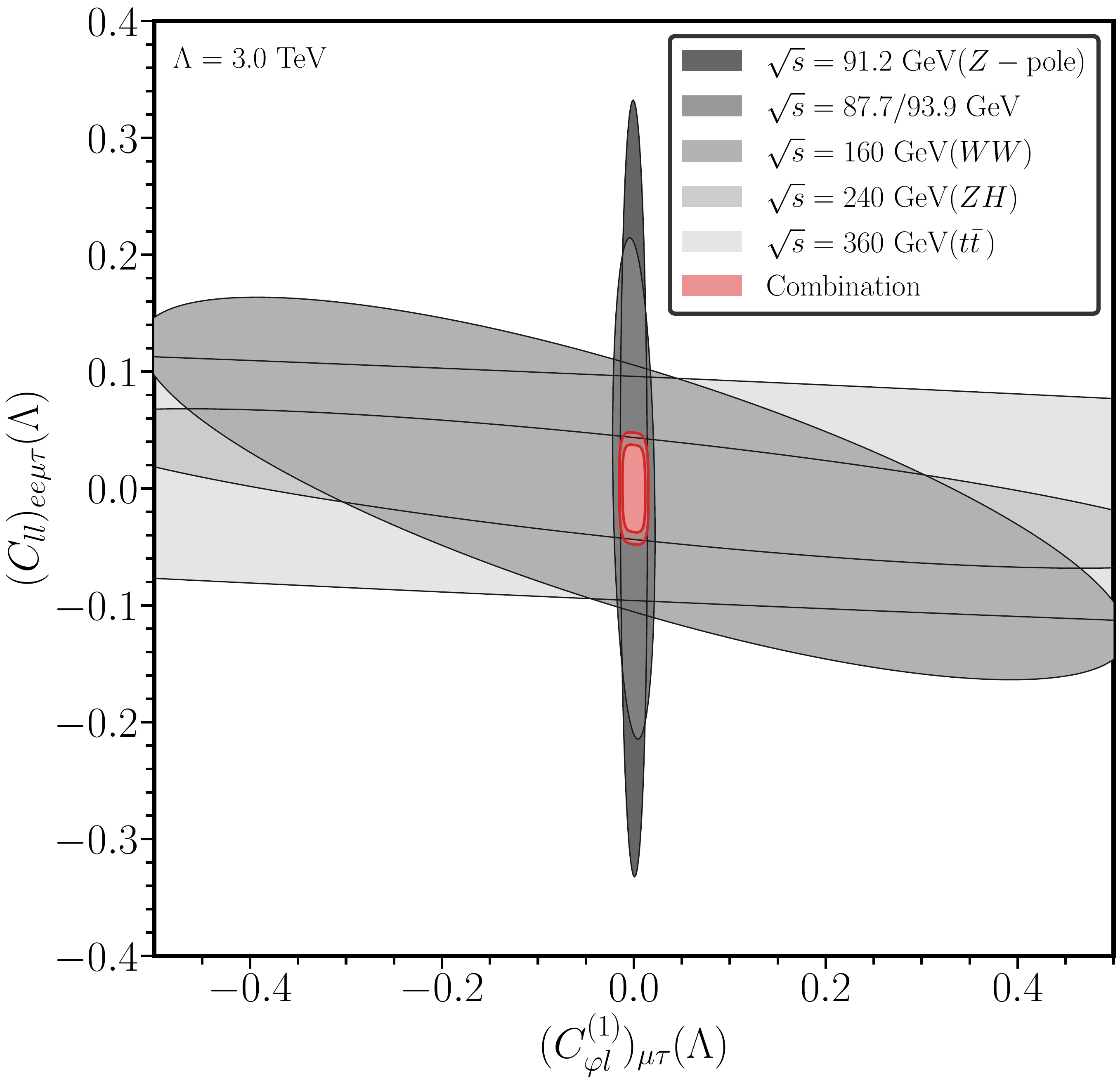}
\caption{$2\sigma$ sensitivity constraints in the plane of the SMEFT Wilson coefficients $(C_{ll})_{ee\mu\tau}$ vs. $(C_{\varphi l}^{(1)})_{\mu\tau}$ (left/right : FCC-ee/CEPC) with a UV scale of $\Lambda=3$ TeV. The red region represents the combined constraints using a $\chi^2$ statistic.} 
\label{fig:Wilson_complementary_1}
\end{figure}

\begin{figure}[!b]
\centering
\includegraphics[width=0.46\linewidth]{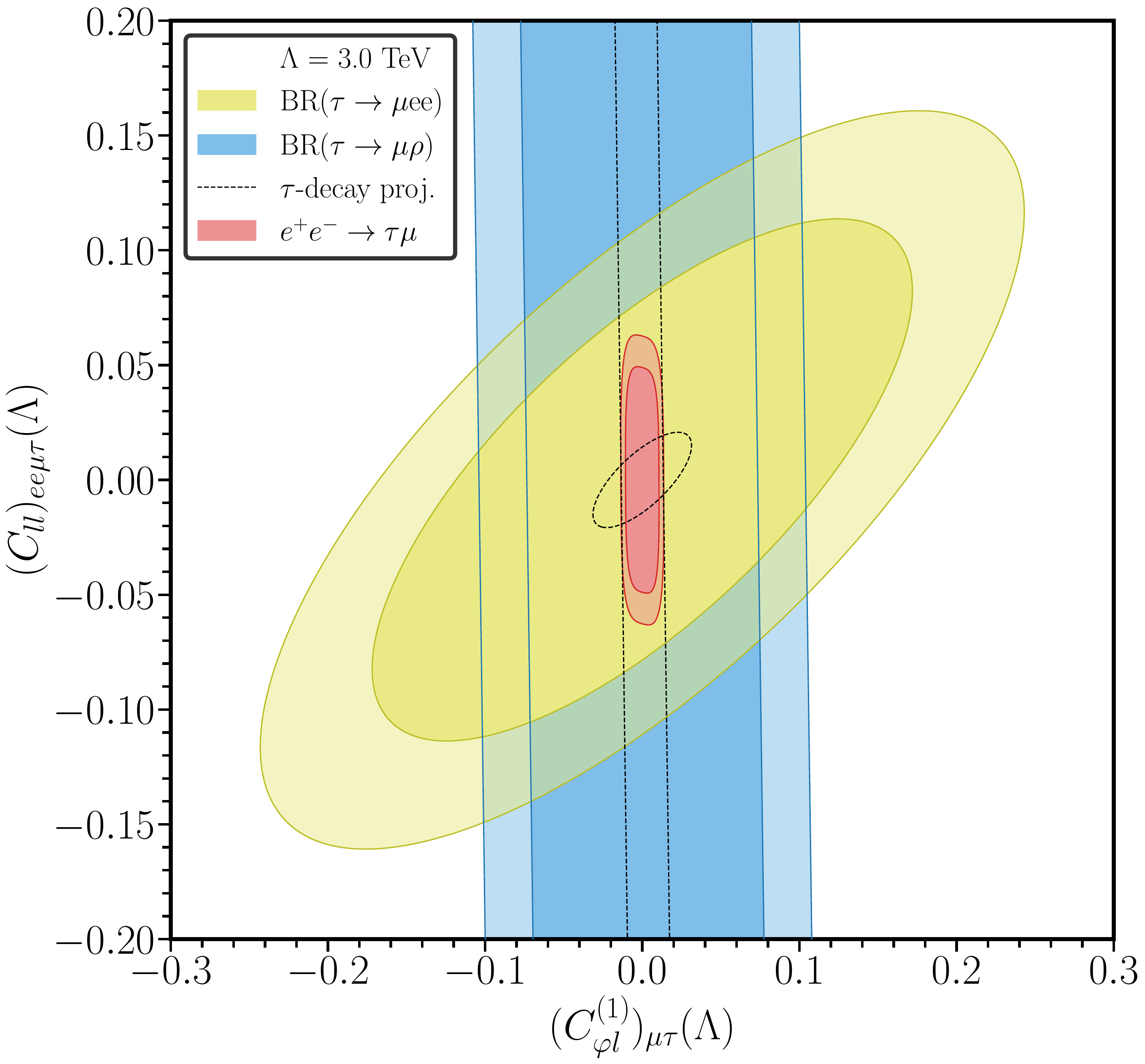} ~~~~
\includegraphics[width=0.46\linewidth]{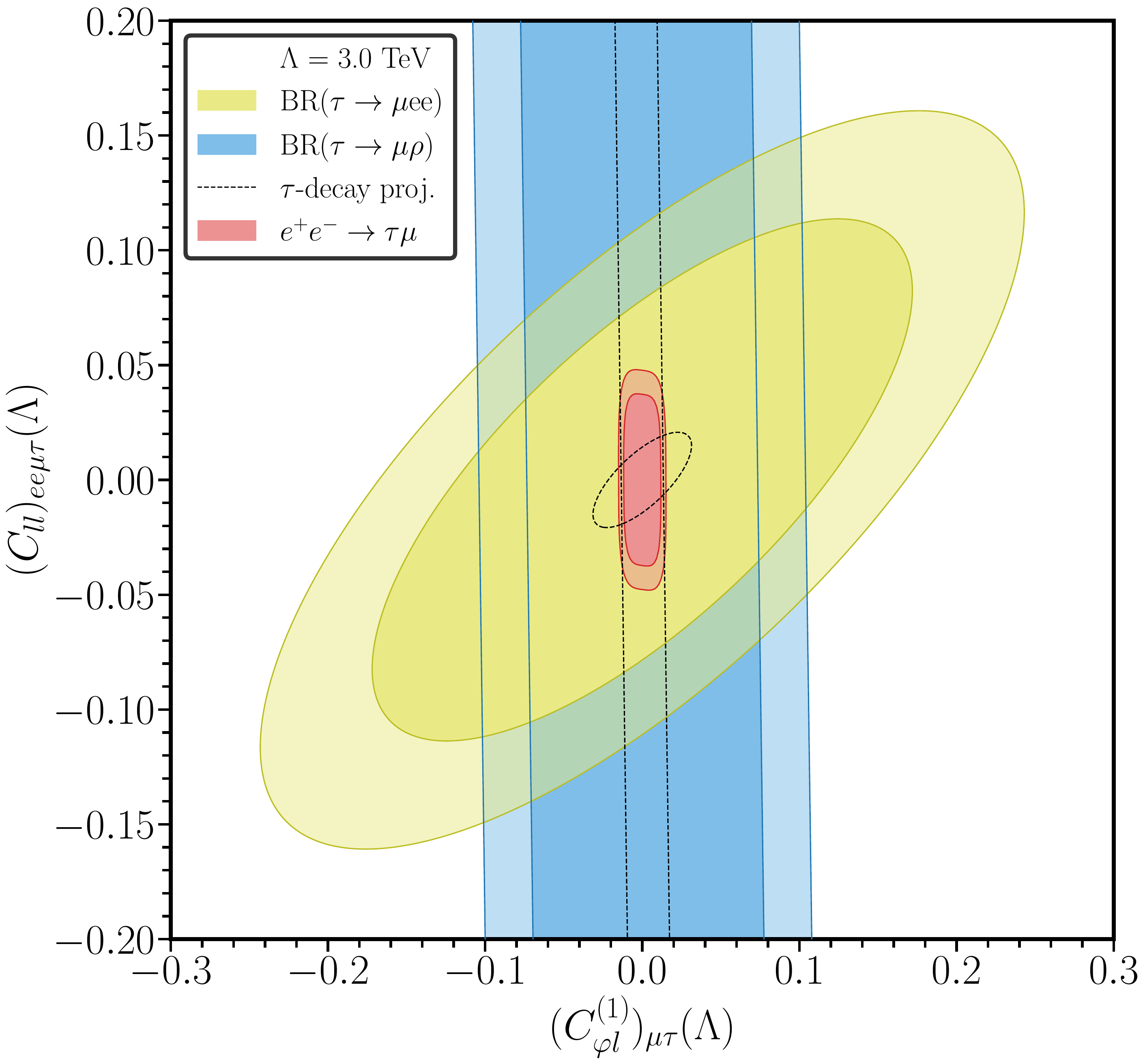}
\caption{Constraints in the plane of the SMEFT Wilson coefficients $(C_{ll})_{ee\mu\tau}$ vs. $(C_{\varphi l}^{(1)})_{\mu\tau}$. The red region shows the combined constraint at FCC-ee (left) and CEPC (right) from figure~\ref{fig:Wilson_complementary_1}.
The yellow (blue) region is the current $1\sigma$ and $2\sigma$ constraints from $\tau \to \mu ee$ ($\tau \to \mu \rho$). The black dashed lines show the expected tau decay constraints at $2\sigma$ from Belle II.} 
\label{fig:Wilson_complementary_2}
\end{figure}

\begin{figure}[!htb]
\centering
\includegraphics[width = \linewidth]{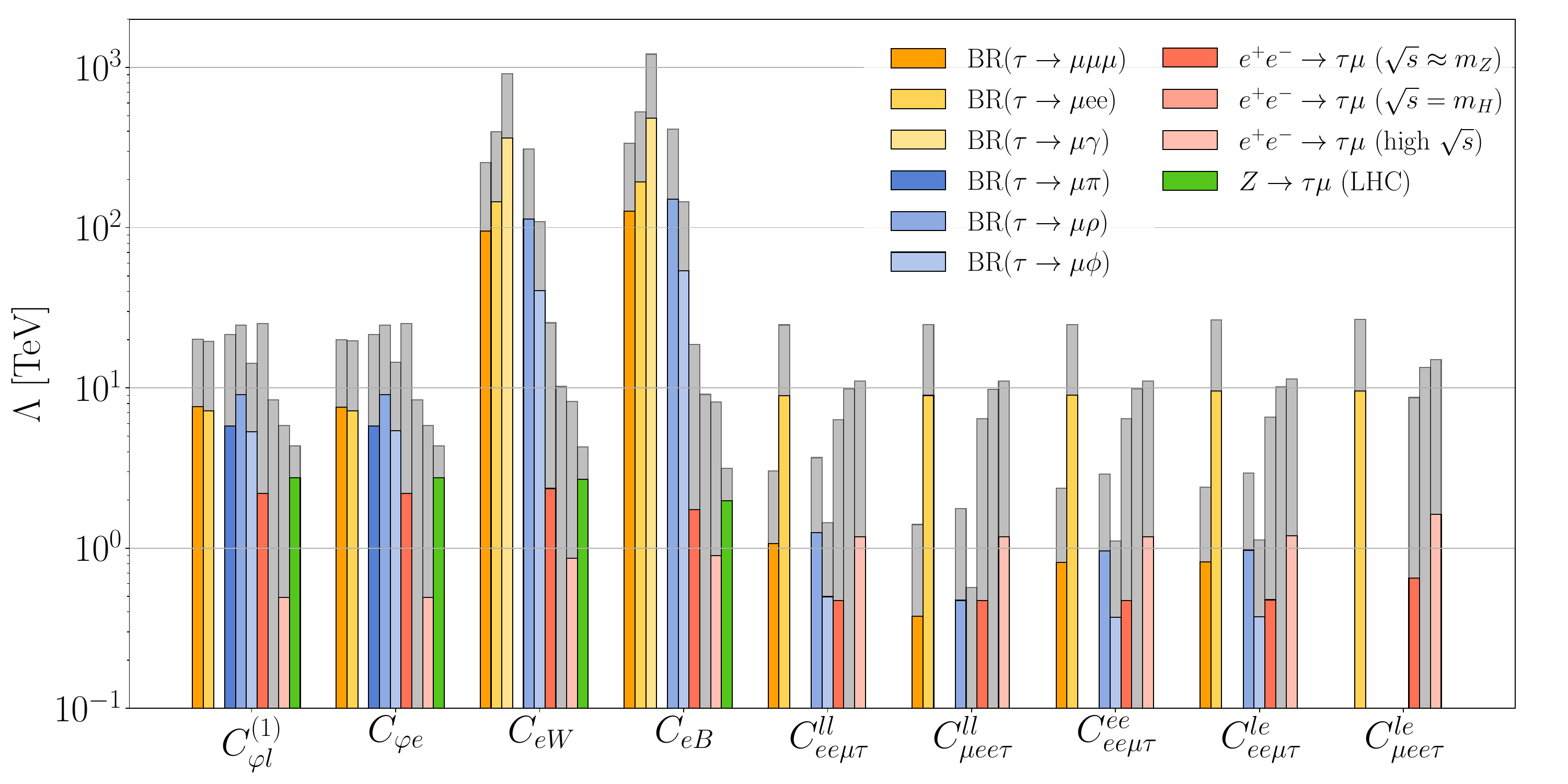}
\caption{Sensitivity to the new physics scale $\Lambda$ from LFV tau decays, $Z$ decays, and the $e^+e^- \to \tau \mu$ cross-section when each SMEFT coefficient is individually set to 1 at the scale $\Lambda$, i.e. $C_i(\Lambda)=1.0$. Current constraints from Babar~\cite{BaBar:2006jhm}, Belle~\cite{Belle:2011ogy, Belle:2021ysv, Hayasaka:2010np}, LEP and the LHC~\cite{ATLAS:2020zlz, ATLAS:2021bdj} are shown as colored bars while future projections from Belle-II, FCC-ee and the HL-LHC~\cite{Altmannshofer:2022fvz, BejarAlonso2020} are shown as gray bars.} 
\label{fig:bar_chart}
\end{figure}

The results are shown in Figures \ref{fig:Wilson_complementary_1} - \ref{fig:bar_chart}. In Figure \ref{fig:Wilson_complementary_1}, it is clear that the four-fermion contact interaction $(C_{ll})_{ee\mu\tau}$ is better constrained at higher $\sqrt{s}$, but that on the $Z$-pole we are much more sensitive to the Higgs current coefficient $(C_{\varphi l}^{(1)})_{\mu\tau}$ than the four-fermion operator. In Figure \ref{fig:Wilson_complementary_2}, we find that the constraints we obtain in the plane of $(C_{\varphi l}^{(1)})_{\mu\tau}$ vs $(C_{ll})_{ee\mu\tau}$ complement the projected constraints of the low-energy tau decays $\tau\to\mu e e $ and $\tau\to\mu\rho$ from Belle-II~\cite{Belle-II:2018jsg}. Finally, in Figure \ref{fig:bar_chart}, we observe that for Higgs current operators and four-fermion contact operators, the FCC-ee and CEPC will be able to probe UV scales $\Lambda \gtrsim O(10)$ TeV rivaling the sensitivity projections from Belle-II.

\section{Conclusion}

We showed that LFV in the process $e^+e^-\to\tau\mu$ would allow us to probe lepton flavor violating dipole operators, Higgs-current operators and four-fermion operators. In particular, FCC-ee and CEPC will have sensitivities that are competitive with future $\tau$-decay constraints from Belle-II. 

Linear colliders such as the ILC~\cite{ILC:2013jhg} and CLIC~\cite{CLICdp:2018cto, Brunner:2022usy} are expected to provide much more stringent bounds on the four fermion operators due to their cross-section linearly scaling with $s$. Furthermore, since these linear colliders typically employ polarized $e^-/e^+$ beams, they will be able to probe the chirality structure of the operators. A detailed study of LFV at future linear $e^+/e^-$ colliders will appear soon~\cite{Altmannshofer:2024}.

\clearpage

\section*{Acknowledgements}
Work presented in these proceedings is based on Ref.~\cite{Altmannshofer:2023tsa}. P. M. thanks Wolfgang Altmannshofer for helpful comments. P. M. also thanks Swagato Banerjee and all the other organizers of the conference for the invitation to present this work.
The research of P. M. is supported by the U.S. Department of Energy grant number DE-SC0010107.

\begin{appendix}

\section{Coefficients}
The Wilson coefficients appearing in the expressions below are linear combinations of the SMEFT coefficients defined in \cite{Grzadkowski:2010es}, see \cite{Altmannshofer:2023tsa}.
\label{app:coeff}
\begin{multline}
I_0 + \bar I_0 = \frac{s}{2 m_Z^2} \left( |C_V^{LL}|^2 + |C_V^{RR}|^2 + |C_V^{LR}|^2 + |C_V^{RL}|^2 + \frac{1}{2}|C_S^{LR}|^2 + \frac{1}{2}|C_S^{RL}|^2 \right) \\
+ \frac{s^2}{(s - m_Z^2)^2 + \Gamma_Z^2 m_Z^2} \Bigg[ \frac{m_Z^2}{2 s} \Big( |C_Z^{LL}|^2 + |C_Z^{RR}|^2 \Big)(1 - 4 s_W^2 + 8 s_W^4) + \left( 1 - \frac{m_Z^2}{s} \right) \\
\times  \left( \text{Re}\Big( C_V^{LL} C_Z^{LL*} + C_V^{LR} C_Z^{RR*}\Big) (1 - 2 s_W^2) - \text{Re}\Big( C_V^{RL} C_Z^{LL*} + C_V^{RR} C_Z^{RR*}\Big) 2 s_W^2 \right) \Bigg] ~,
\end{multline}

\begin{multline}
I_2 + \bar I_2 = 8 \left( |C_\gamma^{LR}|^2 + |C_\gamma^{RL}|^2\right) c_W^2 s_W^2 + \frac{s}{4 m_Z^2} \left( |C_S^{LR}|^2 + |C_S^{RL}|^2\right) \\
+ \frac{s^2}{(s - m_Z^2)^2 + \Gamma_Z^2 m_Z^2} \Bigg[ \Big( |C_Z^{LR}|^2 + |C_Z^{RL}|^2 \Big)(1 - 4 s_W^2 + 8 s_W^4) \\ 
+ \left( 1 - \frac{m_Z^2}{s} \right) \text{Re}\Big( C_\gamma^{LR} C_Z^{LR*} + C_\gamma^{RL} C_Z^{RL*}\Big) 4 c_W s_W (1 - 4 s_W^2)  \Bigg] ~.
\end{multline}

\end{appendix}

\bibliography{bibliography.bib}

\nolinenumbers

\end{document}